\documentclass[aps,prl,reprint,twocolumn,nopacs,superscriptaddress]{revtex4}
\usepackage{amsmath}
\usepackage{amssymb}
\usepackage{graphicx}
\usepackage{hyperref}
\newcommand{\beq}{\begin{equation*}}
\newcommand{\eeq}{\end{equation*}}
\newcommand{\comp}{Mo$_x$W$_{1-x}$Te$_2$}
\newcommand{\half}{Mo$_{0.45}$W$_{0.55}$Te$_2$}

\newcommand{\forty}{Mo$_{0.4}$W$_{0.6}$Te$_2$}

\begin{document}

\title{Measuring Chern numbers above the Fermi level in the Type II\\ Weyl semimetal \comp}

\author{Ilya Belopolski\footnote{These authors contributed equally to this work.}} \email{ilyab@princeton.edu}
\affiliation{Laboratory for Topological Quantum Matter and Spectroscopy (B7), Department of Physics, Princeton University, Princeton, New Jersey 08544, USA}

\author{Su-Yang Xu$^*$}
\affiliation{Laboratory for Topological Quantum Matter and Spectroscopy (B7), Department of Physics, Princeton University, Princeton, New Jersey 08544, USA}

\author{Yukiaki Ishida$^*$}
\affiliation{The Institute for Solid State Physics (ISSP), University of Tokyo, Kashiwa-no-ha, Kashiwa, Chiba 277-8581, Japan}

\author{Xingchen Pan$^*$}
\affiliation{National Laboratory of Solid State Microstructures, Collaborative Innovation Center of Advanced Microstructures, and Department of Physics, Nanjing University, Nanjing, 210093, P. R. China}

\author{Peng Yu$^*$}
\affiliation{Centre for Programmable Materials, School of Materials Science and Engineering, Nanyang Technological University, 639798, Singapore}

\author{Daniel S. Sanchez}
\affiliation{Laboratory for Topological Quantum Matter and Spectroscopy (B7), Department of Physics, Princeton University, Princeton, New Jersey 08544, USA}

\author{Madhab Neupane}
\affiliation{Department of Physics, University of Central Florida, Orlando, FL 32816, USA}

\author{Nasser Alidoust}
\affiliation{Laboratory for Topological Quantum Matter and Spectroscopy (B7), Department of Physics, Princeton University, Princeton, New Jersey 08544, USA}

\author{Guoqing Chang}
\affiliation{Centre for Advanced 2D Materials and Graphene Research Centre, National University of Singapore, 6 Science Drive 2, 117546, Singapore}
\affiliation{Department of Physics, National University of Singapore, 2 Science Drive 3, 117546, Singapore}

\author{Tay-Rong Chang}
\affiliation{Department of Physics, National Tsing Hua University, Hsinchu 30013, Taiwan}

\author{Yun Wu}
\affiliation{Ames Laboratory, U.S. DOE and Department of Physics and Astronomy, Iowa State University, Ames, Iowa 50011, USA}

\author{Guang Bian}
\affiliation{Laboratory for Topological Quantum Matter and Spectroscopy (B7), Department of Physics, Princeton University, Princeton, New Jersey 08544, USA}

\author{Hao Zheng}
\affiliation{Laboratory for Topological Quantum Matter and Spectroscopy (B7), Department of Physics, Princeton University, Princeton, New Jersey 08544, USA}

\author{Shin-Ming Huang}
\affiliation{Centre for Advanced 2D Materials and Graphene Research Centre, National University of Singapore, 6 Science Drive 2, 117546, Singapore}
\affiliation{Department of Physics, National University of Singapore, 2 Science Drive 3, 117546, Singapore}
\affiliation{Department of Physics, National Sun Yat-sen University, Kaohsiung 80424, Taiwan}

\author{Chi-Cheng Lee}
\affiliation{Centre for Advanced 2D Materials and Graphene Research Centre, National University of Singapore, 6 Science Drive 2, 117546, Singapore}
\affiliation{Department of Physics, National University of Singapore, 2 Science Drive 3, 117546, Singapore}

\author{Daixiang Mou}
\affiliation{Ames Laboratory, U.S. DOE and Department of Physics and Astronomy, Iowa State University, Ames, Iowa 50011, USA}

\author{Lunan Huang}
\affiliation{Ames Laboratory, U.S. DOE and Department of Physics and Astronomy, Iowa State University, Ames, Iowa 50011, USA}

\author{You Song}
\affiliation{State Key Laboratory of Coordination Chemistry, School of Chemistry and Chemical Engineering, Collaborative Innovation Center of Advanced Microstructures,�Nanjing University, Nanjing, 210093, P. R. China}

\author{Baigeng Wang}
\affiliation{National Laboratory of Solid State Microstructures, Collaborative Innovation Center of Advanced Microstructures, and Department of Physics, Nanjing University, Nanjing, 210093, P. R. China}

\author{Guanghou Wang}
\affiliation{National Laboratory of Solid State Microstructures, Collaborative Innovation Center of Advanced Microstructures, and Department of Physics, Nanjing University, Nanjing, 210093, P. R. China}

\author{Yao-Wen Yeh}
\affiliation{Princeton Institute for Science and Technology of Materials, Princeton University, Princeton, New Jersey, 08544, USA}

\author{Nan Yao}
\affiliation{Princeton Institute for Science and Technology of Materials, Princeton University, Princeton, New Jersey, 08544, USA}

\author{Julien E. Rault}
\affiliation{Synchrotron SOLEIL, L'Orme des Merisiers, Saint-Aubin-BP 48, 91192 Gif-sur-Yvette, France}

\author{Patrick Le F\`evre}
\affiliation{Synchrotron SOLEIL, L'Orme des Merisiers, Saint-Aubin-BP 48, 91192 Gif-sur-Yvette, France}

\author{Fran\c{c}ois Bertran}
\affiliation{Synchrotron SOLEIL, L'Orme des Merisiers, Saint-Aubin-BP 48, 91192 Gif-sur-Yvette, France}

\author{Horng-Tay Jeng}
\affiliation{Department of Physics, National Tsing Hua University, Hsinchu 30013, Taiwan}
\affiliation{Institute of Physics, Academia Sinica, Taipei 11529, Taiwan}

\author{Takeshi Kondo}
\affiliation{The Institute for Solid State Physics (ISSP), University of Tokyo, Kashiwa-no-ha, Kashiwa, Chiba 277-8581, Japan}

\author{Adam Kaminski}
\affiliation{Ames Laboratory, U.S. DOE and Department of Physics and Astronomy, Iowa State University, Ames, Iowa 50011, USA}

\author{Hsin Lin}
\affiliation{Centre for Advanced 2D Materials and Graphene Research Centre, National University of Singapore, 6 Science Drive 2, 117546, Singapore} \affiliation{Department of Physics, National University of Singapore, 2 Science Drive 3, 117546, Singapore}

\author{Zheng Liu} \email{z.liu@ntu.edu.sg}
\affiliation{Centre for Programmable Materials, School of Materials Science and Engineering, Nanyang Technological University, 639798, Singapore}
\affiliation{NOVITAS, Nanoelectronics Centre of Excellence, School of Electrical and Electronic Engineering, Nanyang Technological University, 639798, Singapore}
\affiliation{CINTRA CNRS/NTU/THALES, UMI 3288, Research Techno Plaza, 50 Nanyang Drive, Border X Block, Level 6, 637553, Singapore}

\author{Fengqi Song} \email{songfengqi@nju.edu.cn}
\affiliation{National Laboratory of Solid State Microstructures, Collaborative Innovation Center of Advanced Microstructures, and Department of Physics, Nanjing University, Nanjing, 210093, P. R. China}

\author{Shik Shin}
\affiliation{The Institute for Solid State Physics (ISSP), University of Tokyo, Kashiwa-no-ha, Kashiwa, Chiba 277-8581, Japan}

\author{M. Zahid Hasan} \email{mzhasan@princeton.edu}
\affiliation{Laboratory for Topological Quantum Matter and Spectroscopy (B7), Department of Physics, Princeton University, Princeton, New Jersey 08544, USA}
\affiliation{Princeton Institute for Science and Technology of Materials, Princeton University, Princeton, New Jersey, 08544, USA}

\pacs{}

\begin{abstract}
It has recently been proposed that electronic band structures in crystals give rise to a previously overlooked type of Weyl fermion, which violates Lorentz invariance and, consequently, is forbidden in particle physics. It was further predicted that \comp\ may realize such a Type II Weyl fermion. One crucial challenge is that the Weyl points in \comp\ are predicted to lie above the Fermi level. Here, by studying a simple model for a Type II Weyl cone, we clarify the importance of accessing the unoccupied band structure to demonstrate that \comp\ is a Weyl semimetal. Then, we use pump-probe angle-resolved photoemission spectroscopy (pump-probe ARPES) to directly observe the unoccupied band structure of \comp. For the first time, we directly access states $> 0.2$ eV above the Fermi level. By comparing our results with \textit{ab initio} calculations, we conclude that we directly observe the surface state containing the topological Fermi arc. Our work opens the way to studying the unoccupied band structure as well as the time-domain relaxation dynamics of \comp\ and related transition metal dichalcogenides.
\end{abstract}

\date{\today}
\maketitle

\begin{figure}
\centering
\includegraphics[width=8.5cm, trim={0 0 0 0}, clip]{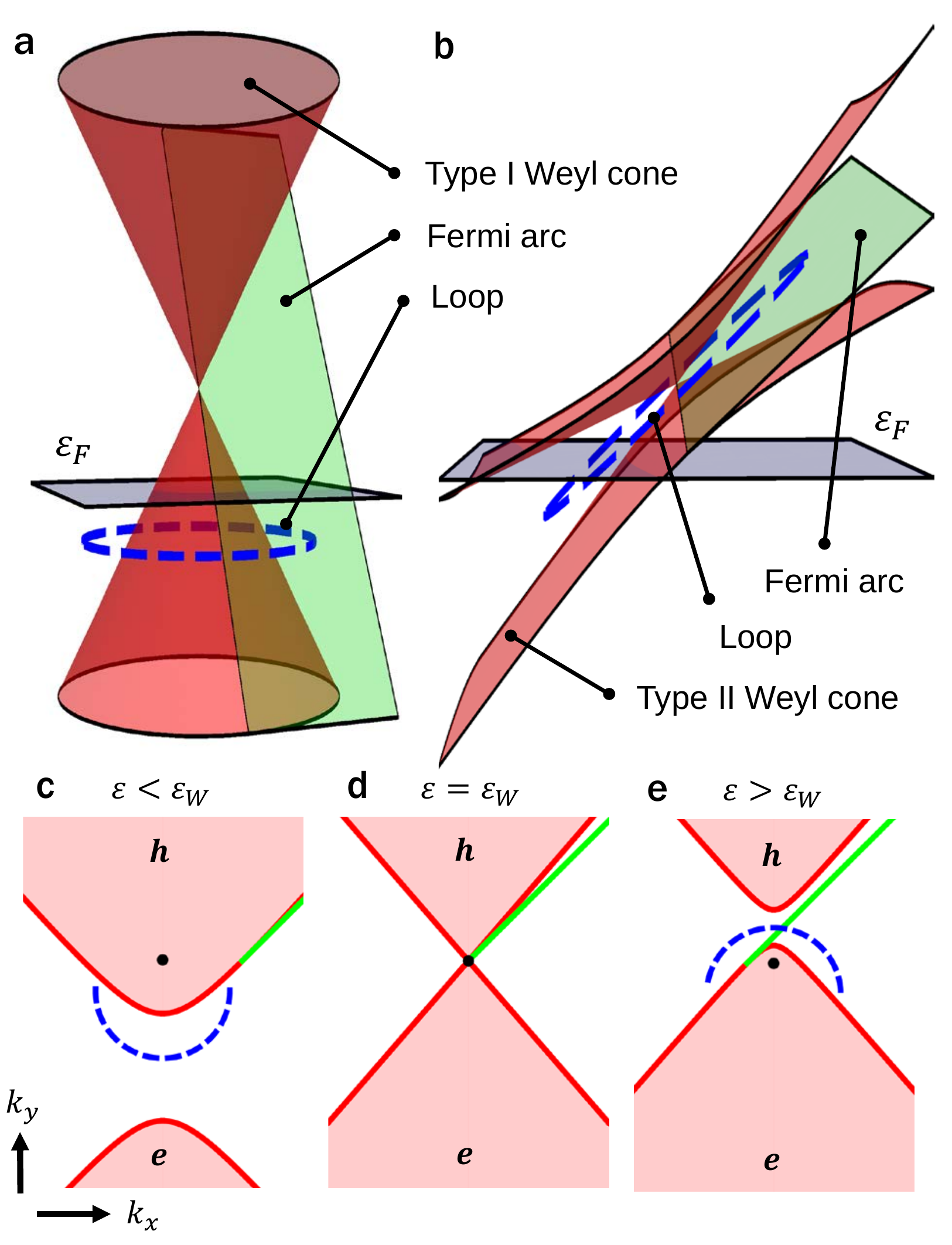}
\caption{\label{Fig1}\textbf{Chern numbers in Type II Weyl semimetals.} A loop (dashed blue line) to show a nonzero Chern number in (a) a Type I Weyl cone and (b) a Type II Weyl cone. Suppose the Weyl point lies above $\varepsilon_F$. In the simplest case, it is clear that for a Type I Weyl cone, we can show a nonzero Chern number by counting crossings of surface states on a closed loop which lies entirely below the Fermi level. This is not true for a Type II Weyl semimetal. (c-e) Constant-energy cuts of (b). We can attempt to draw a loop below $\varepsilon_F$, as in (c), but we find that the loop runs into the bulk hole pocket. We can close the loop by going to $\varepsilon > \varepsilon_W$, as in (e). However, then the loop must extend above $\varepsilon_F$.}
\end{figure}

\begin{figure*}
\centering
\includegraphics[width=15cm, trim={0 0 0 0}, clip]{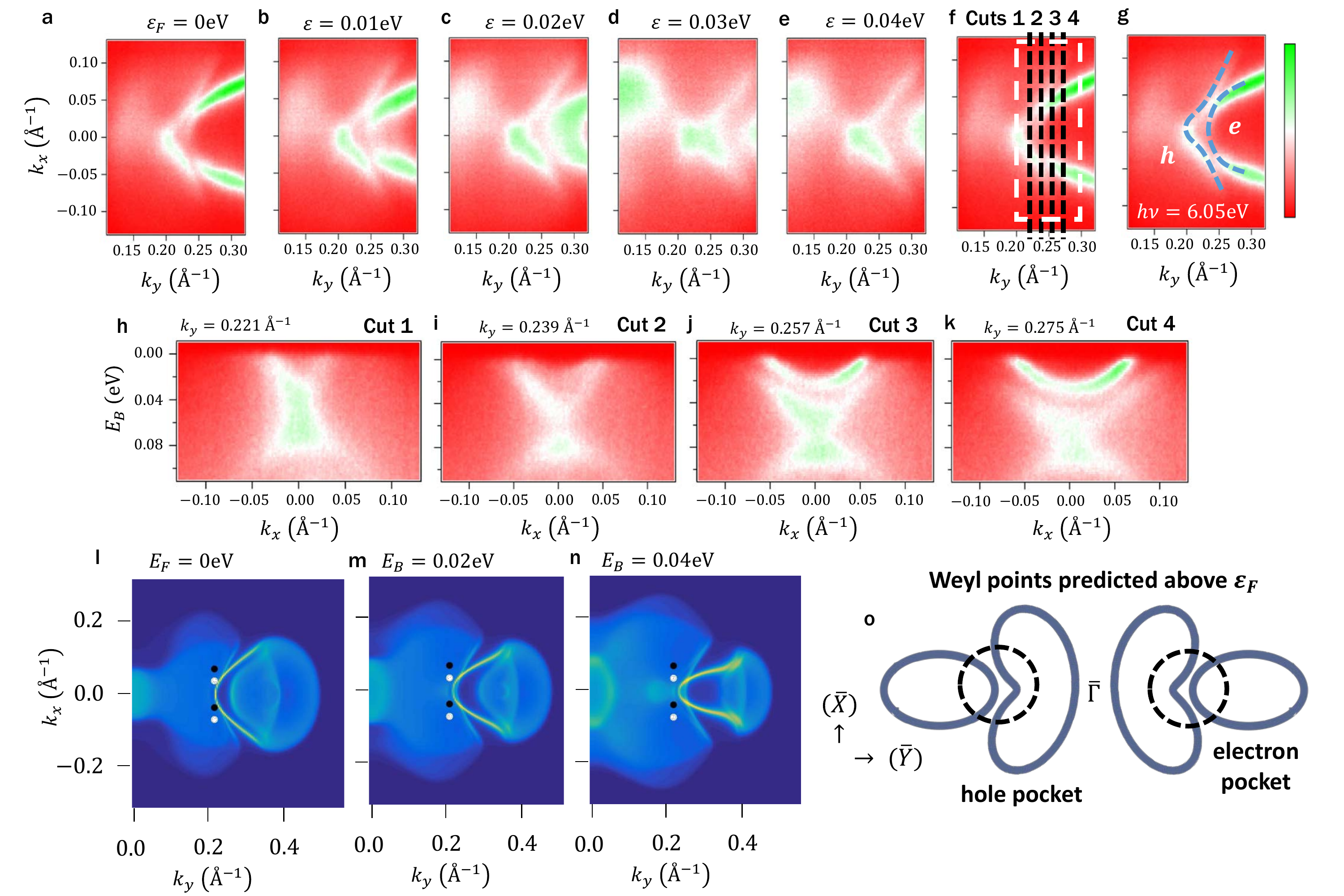}
\caption{\label{Fig2}\textbf{\half\ below the Fermi level.} (a-e) Conventional laser ARPES spectra of the constant-energy surface at different binding energies using incident photon energy $h\nu=6.05$ eV. We observe a palmier-shaped hole pocket and an almond-shaped electron pocket, see also the Supplemental Material \cite{SM}. The two pockets approach each other and we directly observe a beautiful avoided crossing near $\varepsilon_F$ where they hybridize, in (a). This hybridization is expected to give rise to Weyl point above the Fermi level. (f) Same data as (a) with black dotted lines showing the locations of $E_{\textrm{B}}-k_y$ dispersion maps and a white dotted box showing the momentum-space range of a Fermi surface mapping above the Fermi level, shown in Fig. \ref{Fig3}. (g) The same data as panel (a) with blue dotted lines that roughly trace the electron, $e$, and hole, $h$, pockets at the Fermi level. (h-k) ARPES measured $E_{\textrm{B}}-k_y$ dispersion maps along the cuts shown in (f). The electron and hole pockets nest into each other. We expect Weyl points or Fermi arcs above the Fermi level at certain $k_y$ where the pockets approach. (l-n) Constant energy contours for \forty\ from \textit{ab initio} calculations, showing clearly the palmier-shaped hole pocket and almond-shaped electron pocket. The black and white dots indicate the locations of the Weyl points, above $\varepsilon_F$. Note the excellent overall agreement with calculation. The offset on the $k_y$ scale on the ARPES spectra is set by comparison with calculation. (o) Cartoon of the palmier and almond at the Fermi level. Based on \textit{ab initio} calculation, we expect Weyl points above $\varepsilon_F$ where the pockets intersect.}
\end{figure*}

\begin{figure*}
\centering
\includegraphics[width=15cm, trim={10 10 10 10}, clip]{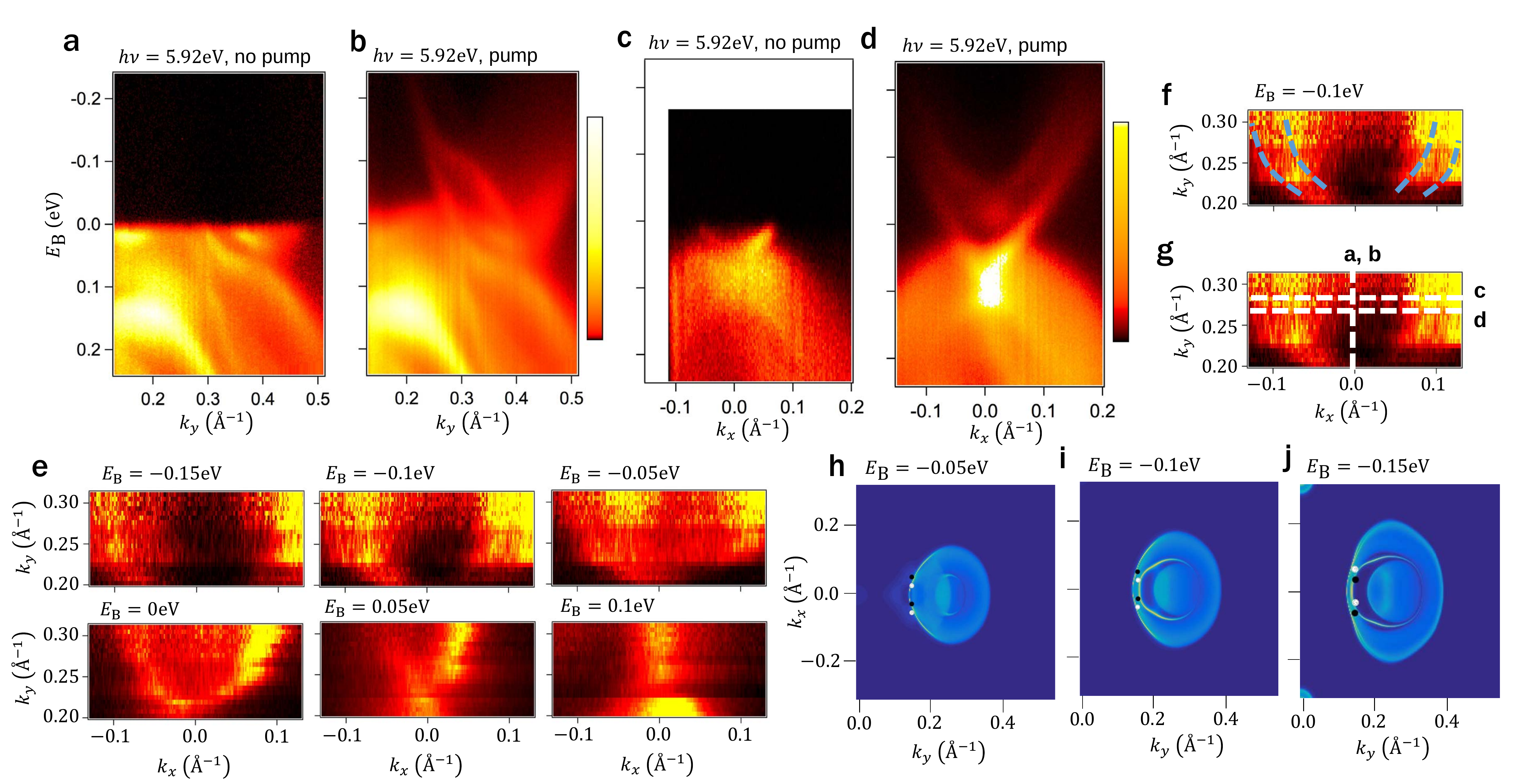}
\caption{\label{Fig3}\textbf{\half\ above the Fermi level.} (a,b) $E_{\textrm{B}}-k_y$ dispersion maps of \half\ along the $\bar{\Gamma}$-$\bar{Y}$ direction at $k_{x} = 0$ with and without the pump laser. The sample responds beautifully to the pump laser, allowing us access to the band structure $ > 0.2$eV above $E_{F}$, well above the predicted energies of the Weyl points. (c,d) Dispersion maps of \half\ with and without pump on a cut parallel to $\bar{\Gamma}$-$\bar{X}$ at (c) $k_{y}\sim 0.29 \textrm{\AA}$ and (d) $k_{y}\sim 0.26 \textrm{\AA}$. Without pump we see a single electron band, associated with the almond pocket. With pump, we find an additional electron band with minimum at $\sim -0.02$eV. We also observe a discontinuity in this band at $\sim \pm 0.02 \textrm{\AA}^{-1}$. (e) The evolution of the Fermi surface for the almond pocket in energy. We observe that the almond pocket evolves into two nested contours, seen most clearly at $E_{B}=-0.1$eV. (f) Same data as (e) at $E_B = -0.1$eV, with dotted blue guides to the eye to mark the two nested contours. (g) Cuts shown in (a-d) labeled for reference. (h-j) Calculation of the Fermi surface for \forty\ above $E_F$. We see two nested electron pockets, consistent with the measured Fermi surface at $E_B = -0.1$eV.}
\end{figure*}

Weyl fermions have been known since the early twentieth century as chiral particles associated with solutions to the Dirac equation at zero mass \cite{Weyl, Peskin}. In particle physics, imposing Lorentz invariance uniquely fixes the dispersion for a Weyl fermion. However, effective field theories in condensed matter physics are not required to obey Lorentz invariance, leaving a freedom in the Weyl fermion dispersion. Recently, it was discovered that this freedom allows a new type of Weyl fermion to arise in a crystalline band structure, distinct from the Weyl fermion relevant to particle physics \cite{AndreiNature, Grushin, Bergholtz, Trescher, Beenakker, Zyuzin, Isobe}. This Type II Weyl fermion strongly violates Lorentz invariance and has a dispersion characterized by a Weyl cone which is tilted over on its side. It was further predicted that a Type II Weyl semimetal arises in WTe$_2$ \cite{AndreiNature}. Concurrently, MoTe$_2$ and \comp\ were predicted to be Weyl semimetals \cite{TayRong, Binghai, Zhijun} and, more recently, several additional Type II Weyl semimetal candidates have been proposed \cite{Ta3S2, Koepernik, Autes}. All theoretical studies found that all Weyl points in the \comp\ series are above the Fermi level, $\varepsilon_F$. While angle-resolved photoemission spectroscopy (ARPES) would be the technique of choice to directly demonstrate a Type II Weyl semimetal in \comp, conventional ARPES can only study occupied electron states, below $\varepsilon_F$, making it challenging to access the Weyl semimetal state in \comp. Nonetheless, several ARPES works provide evidence of a Type II Weyl semimetal in MoTe$_2$ and WTe$_2$ by studying the band structure above the Fermi level in the tail of the Fermi-Dirac distribution \cite{Adam1, Adam2, Xinjiang2}, while other works have attempted to demonstrate a Type II Weyl semimetal in MoTe$_2$ and WTe$_2$ in ARPES by studying only the band structure below $\varepsilon_F$ \cite{Shuyun, Chen, Xinjiang, HongDing, Baumberger}. We note also a recent experimental study of Type II Weyl fermions in an unrelated compound \cite{LaAlGe, RAlX}.

Here, we first argue that it's crucial to study the band structure above $\varepsilon_F$ to show a Weyl semimetal in \comp, even if the Fermi arcs fall partly below the Fermi level. Then, we experimentally demonstrate that we can access states sufficiently above the Fermi level in \comp\ using a state-of-the-art photoemission technique known as pump-probe ARPES. Our work sets the stage for experimentally demonstrating the first Type II Weyl semimetal in \comp. We also show that we can study the unoccupied band structure and time-relaxation dynamics of related transition metal dichalcogenides using pump-probe ARPES.

Can we show that a material is a Weyl semimetal if the Weyl points are above the Fermi level, $\varepsilon_W > \varepsilon_F$? For the simple case of a well-separated Type I Weyl point of chiral charge $\pm 1$, it is easy to see that this is true. Specifically, although we cannot see the Weyl point itself, the Fermi arc extends below the Fermi level, see Fig. \ref{Fig1}(a). Therefore, we can consider a closed loop in the surface Brillouin zone which encloses the Weyl point. By counting the number of surface state crossings on this loop we can demonstrate a nonzero Chern number \cite{NbPme}. In our example, we expect one crossing along the loop. By contrast, this approach fails for a well-separated Type II Weyl point above $\varepsilon_F$. In particular, recall that when counting Chern numbers, the loop we choose must stay always in the bulk band gap. As a result, for the Type II case we cannot choose the same loop as in the Type I case since the loop would run into the bulk hole pocket. We might instead choose a loop which is slanted in energy, see Fig. \ref{Fig1}(b), but such a loop would necessarily extend above $\varepsilon_F$. Alternatively, we can consider different constant energy cuts of the Type II Weyl cone. In Figs. \ref{Fig1}(c-e), we show constant-energy cuts of the Type II Weyl cone and Fermi arc. We see that we cannot choose a closed loop around the Weyl point by looking only at one energy, because the loop runs into a bulk pocket. However, we can build up a closed loop from segments at energies above the below the Weyl point, as in Figs. \ref{Fig1}(c,e). But again, we must necessarily include a segment on a cut at $\varepsilon > \varepsilon_W$. We find that for a Type II Weyl semimetal, if the Weyl points are above the Fermi level, we must study the unoccupied band structure.

Next, we argue that in the specific case of \comp\ we must access the unoccupied band structure to show a Weyl semimetal. In the Supplemental Material, we present a detailed discussion of the band structure of \comp\  \cite{SM}. Here, we only note a key result, consistent among all \textit{ab initio} calculations of \comp, that all Weyl points are Type II and are above the Fermi level \cite{AndreiNature, TayRong, Zhijun, Binghai}. These facts are essentially sufficient to require that we access the unoccupied band structure. However, it is useful to provide a few more details. Suppose that the Fermi level of \comp\ roughly corresponds to the case of Fig. \ref{Fig1}(c). To count a Chern number using only the band structure below the Fermi level, we need to find a path enclosing a nonzero chiral charge while avoiding the bulk hole and electron pockets. We can try to trace a path around the entire hole pocket. However, as we will see in Fig. \ref{Fig2}, in \comp, the Weyl point projections all fall in one large hole pocket at $\varepsilon_F$. As a result, tracing around the entire hole pocket encloses zero chiral charge. Therefore, demonstrating a Weyl semimetal in \comp\ requires accessing the unoccupied band structure. 

Now, we introduce the occupied band structure of \comp. We present ARPES spectra of the Fermi surface of \half\ below the Fermi level, see Figs. \ref{Fig2}(a-e). At the Fermi level, we observe an electron and a hole pocket which approach each other as we scan the binding energy toward the Fermi level. We also study the evolution of the two pockets on $E_\textrm{B}$-$k_x$ cuts, as indicated in Fig. \ref{Fig2}(f), and we mark the electron and hole pockets in Fig. \ref{Fig2}(g). In the $E_\textrm{B}-k_x$ cuts, shown in Fig.\ref{Fig2}(h-k), we find that the electron and hole pockets approach each other. Based on our \textit{ab initio} calculations, we expect these bands to intersect above $\varepsilon_F$ at $k_y \sim k_\textrm{W}$, forming Weyl points and Fermi arcs. We find excellent agreement between our ARPES results and \textit{ab initio} calculations, see Fig. \ref{Fig2}(l-n). In calculation, we add the ($k_x$, $k_y$) locations of the Weyl points as black and white dots. We also plot a cartoon of the electron and hole pockets at the Fermi level in Fig. \ref{Fig2} (o). We see that as the electron and hole pockets evolve in energy they chase each other. At two energies above $\varepsilon_F$, the pockets catch up to each other, forming two sets of Weyl points $W_1$ and $W_2$. For additional discussion of the configuration of Weyl points, see the Supplemental Material \cite{SM}.

\begin{figure*}
\centering
\includegraphics[width=15cm, trim={0 0 0 0}, clip]{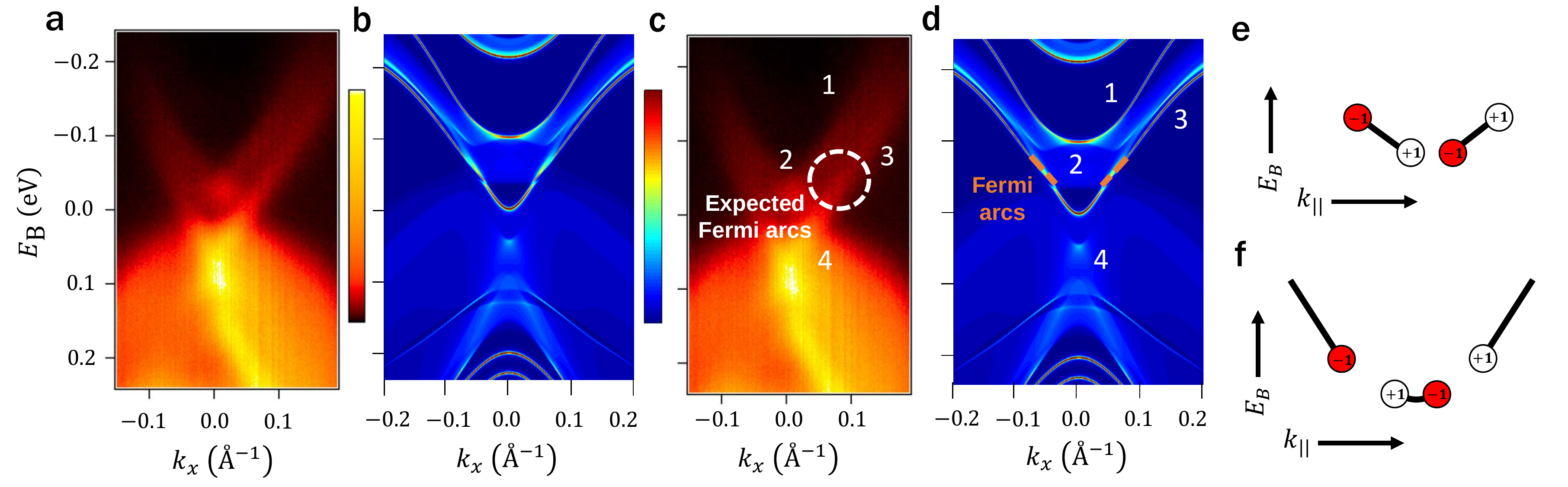}
\caption{\label{Fig4}\textbf{Signatures of Fermi arcs in \comp.} (a) Pump-probe ARPES spectrum at $k_y = 0.225 \textrm{\AA}^{-1}$. (b) \textit{Ab initio} calculation at $k_y = k_\textrm{W1} = 0.215 \textrm{\AA}^{-1}$, showing an excellent overall agreement with the data. (c,d) Same as (a,b) but with key features in the data marked. (e,f) There are two possible scenarios for the connectivity of the arcs. The scenario in (e) is favored by our \textit{ab initio} calculations. The surface state electron pocket (3) contains both the topological Fermi arcs and trivial surface states. These trivial surface states may merge into the bulk bands near the Weyl points, so that we do not expect to observe a disjoint arc. Nonetheless, if the slope of the Fermi arc and trivial surface state is different, it may be possible to observe a ``ripple'' in the surface state (3), providing a direct experimental signature of the Fermi arc. We propose that a future measurement may demonstrate a Weyl semimetal in \comp\  in this way.}
\end{figure*}

Next, we show that we can directly access the unoccupied states in \comp\ with pump-probe ARPES. In our pump-probe ARPES experiment, we use a $1.48$eV pump laser to excite electrons into low-lying states above the Fermi level and a $5.92$eV probe laser to perform photoemission \cite{IshidaMethods}. We first present a cut of \half\ along the $\bar{\Gamma}-\bar{Y}$ direction at $k_x = 0$ with and without the pump laser, see Figs. \ref{Fig3}(a,b). The sample responds beautifully to the pump laser and we observe a dramatic evolution of the bands up to energies $> 0.2$eV above the Fermi level. Note that all available calculations of \comp\ place the Weyl points $< 0.1$eV above the Fermi level. In addition, the Weyl points are all predicted to be within $0.25 \textrm{\AA}^{-1}$ of the $\bar{\Gamma}$ point \cite{AndreiNature, TayRong, Binghai, Zhijun}. We see that our pump-probe measurement easily accesses the relevant region of reciprocal space to show a Weyl semimetal in \comp\ for all $x$. We see in Fig. \ref{Fig3}(b) that the palmier pocket closes up within $\sim -0.02$eV of the Fermi level. However, the almond pocket extends well above the Fermi level. In particular, we notice two branches reaching toward $\bar{\Gamma}$ and appearing to merge at $\sim -0.15$eV. These bands correspond to two distinct surface states corresponding to inner and outer contours of the almond, as seen in calculation. We also see a broader, diffuse branch of the almond pocket reach toward the $\bar{Y}$ point, which may arise from bulk states. Next, we study the band structure on a cut parallel to $\bar{\Gamma}-\bar{X}$ at $k_y \sim 0.36 \textrm{\AA}^{-1}$, with and without the pump laser, see Figs. \ref{Fig3}(c,d). In our pump-probe measurement, we see a pair of electron bands from the almond pocket. The lower branch corresponds to the sharp, prominent branch of the almond as it appears below $E_F$. On the other hand, the upper branch, with a minimum at $\sim -0.02$eV, lies entirely above the Fermi level. We note that this upper branch appears to have a discontinuity at $k_x \sim 0.02 \textrm{\AA}^{-1}$ where it seems to break into two disjoint bands. We comment further on this discontinuity below. To better understand the band structure above the Fermi level, we show a Fermi surface mapping of the almond pocket above $E_F$, see Fig. \ref{Fig3}(e). We see the almond pocket grows larger above $E_F$, again, as an electron pocket. We also find that the almond pocket evolves into two nested contours. This structure is most easily visible at $E_\textrm{B} = -0.1$eV in Fig. \ref{Fig3}(e). We reproduce this constant energy cut in Fig. \ref{Fig3}(f), where we mark the two contours by blue dotted lines. From the $\bar{\Gamma}-\bar{Y}$ cut, we see that this pair of nested contours corresponds to the two contours which make up the almond pocket at $E_F$. For reference, in Fig. \ref{Fig3}(g) we label the cuts shown in Figs. \ref{Fig3}(a-d). We see that our pump-probe ARPES measurements easily access the relevant energy range above the Fermi level where we expect to observe Weyl points and Fermi arcs.

Next, we consider signatures of a Weyl semimetal in \comp\ in pump-probe ARPES. We compare an ARPES cut expected to be near the Weyl points with an \textit{ab initio} calculation for \forty\ at $k_y = k_\textrm{W1}$, see Figs. \ref{Fig4}(a-d). We find excellent overall agreement. Specifically, we find an upper electron pocket (1) with a disjoint segment (2), a lower electron pocket (3) and the approach between hole and electron bands (4). We expect a topological Fermi arc in (3). Since the chiral charges of all Weyl points projections are $\pm 1$, we expect a disjoint arc connecting pairs of Weyl points, in one of two possible configurations, Fig. \ref{Fig4}(e,f). From our calculation, we expect the scenario in Fig. \ref{Fig4}(e), see the dotted orange lines in Fig. \ref{Fig4} (d). However, such a disjoint contour may not necessarily be experimentally relevant. In particular, we expect the large surface state electron pocket (3) to contain both the topological Fermi arc and some additional trivial surface state. Only the Fermi arc is required by the topological invariants to terminate on the Weyl points, but the trivial surface state will also likely merge into the bulk very close to the Weyl points, so that within experimental resolution there is no disjoint contour. We then ask if we can reasonably expect any signature of the Weyl semimetal state in \comp. One possibility is that even if there is no disjoint arc, the slope of the Fermi arc and the trivial surface states are different. In this way, the Weyl points may create a ``ripple'' in the electron pocket (3), providing a direct experimental signature of the Weyl semimetal state. We propose that future measurements may allow us to observe this direct signature of a Weyl semimetal in \half. A composition dependence of this ``ripple'' may further provide conclusive evidence of a Weyl semimetal in \comp.

Despite numerous works (preprints) studying the Weyl semimetal state in \comp\ by ARPES, there has so far only been preliminary evidence for Weyl points and Fermi arcs. Here we demonstrate the importance of studying the bulk or surface band structure above the Fermi level to decisively show a Weyl semimetal in \comp. We also directly access the band structure well above the Fermi level in \half\ using pump-probe ARPES. Our work opens the way to the study of unoccupied band structure and time-domain relaxation dynamics of transition metal dichalcogenides.

\ \\
{\bf Acknowledgments}
\\

I.B. and D.S. thank Moritz Hoesch and Timur Kim for support during synchrotron ARPES measurements at Beamline I05 of Diamond Lightsource in Didcot, UK. I.B. acknowledges the support of the US National Science Foundation GRFP. Y.I. is supported by the Japan Society for the Promotion of Science, KAKENHI 26800165. The ARPES measurements at Ames Lab were supported by the U.S. Department of Energy, Office of Science, Basic Energy Sciences, Materials Science and Engineering Division. Ames Laboratory is operated for the U.S. Department of Energy by Iowa State University under contract No. DE-AC02-07CH11358. M.N. is supported by start-up funds from the University of Central Florida. X.C.P., Y.S., B.G.W., G.H.W and F.Q.S. thank the National Key Projects for Basic Research of China (Grant Nos. 2013CB922100, 2011CB922103), the National Natural Science Foundation of China (Grant Nos. 91421109, 11522432, and 21571097) and the NSF of Jiangsu province (No. BK20130054). This work is also financially supported by the Singapore National Research Foundation (NRF) under NRF RF Award No. NRF-RF2013-08, the start-up funding from Nanyang Technological University (M4081137.070). T.-R.C. and H.-T.J. were supported by the National Science Council, Taiwan. H.-T.J. also thanks the National Center for High-Performance Computing, Computer and Information Network Center National Taiwan University, and National Center for Theoretical Sciences, Taiwan, for technical support. H.L. acknowledges the Singapore NRF under Award No. NRF-NRFF2013-03.

%\pagebreak
\begin{widetext}
\end{widetext}
%\clearpage

\begin{center}
\textbf{\large Supplemental Materials}% for: Measuring Chern numbers above the Fermi level in the Type II Weyl semimetal \comp}
\end{center}

\setcounter{equation}{0}
\setcounter{figure}{0}
\setcounter{table}{0}
\makeatletter
\renewcommand{\theequation}{S\arabic{equation}}
\renewcommand{\thefigure}{S\arabic{figure}}
\renewcommand{\thetable}{S\arabic{table}}

\bigskip
{\bf $\S 1.$ Overview of the crystal structure and electronic structure of \comp}
\bigskip

We briefly present an overview of the system under study. \comp\ crystallizes in an orthorhombic Bravais lattice, space group $Pmn2_1$ ($\#31$), lattice constants $a = 6.282\textrm{\AA}$, $b = 3.496\textrm{\AA}$ and $c = 14.07\textrm{\AA}$ \cite{MoTe2WTe2}. The atomic structure is layered, with single layers of W/Mo sandwiched in between Te bilayers, see Figs. \ref{Fig1}(a,b). The structure has no inversion symmetry, a crucial condition for a Weyl semimetal. Shown in Fig. \ref{Fig1}(c) is a scanning electron microscope (SEM) image of a typical \half\ sample showing a layered crystal structure. Energy-dispersive spectroscopy (EDS) measurements confirm that the crystals have composition \half, see Methods: \half\ growth, below. We display the bulk band structure of WTe$_2$ along high-symmetry lines in Fig. \ref{Fig1}(d), with the bulk and (001) surface Brillouin zone (BZ) of \comp\ shown in Fig. \ref{Fig1}(e). The bulk band structure of WTe$_2$ is gapped throughout the Brillouin zone except near the $\Gamma$ point where the bulk valence and conduction bands approach each other, forming an electron and hole pocket near the Fermi level. Because the crystal structure breaks inversion symmetry, we expect Weyl points to arise generically where the bands hybridize. Although Weyl points have been found in WTe$_2$ in calculation, it is now understood that WTe$_2$ is very near the critical point for a transition to a trivial phase and the Weyl semimetal state is not expected to be robust \cite{TayRong}. However, it has been shown in calculation that an Mo doping causes the bands to further invert, increasing the separation of the Weyl points \cite{TayRong}. This calculation result motivates our study of \comp. Next, we provide an overview of the (001) Fermi surface of \half\ under vacuum ultraviolet ARPES, see Fig. \ref{Fig1}(f,g). We observe the palmier and almond pockets on either side of $\bar{\Gamma}$ along $\bar{\Gamma}-\bar{Y}$. Our results are in excellent overall agreement with \textit{ab initio} calculations of \forty, as shown in Figs. 2(l-n) of the main text. One important discrepancy is the distance between the features and $\bar{\Gamma}$, which is underestimated by $\sim 0.1\textrm{\AA}^{-1}$ in calculation. We speculate that this might be an artifact of the ARPES Fermi surface mapping procedure. Specifically, if the sample tends to form a curved surface \textit{in situ} after cleaving, there may be a drift in the measured slice of momentum space. In Fig. \ref{Fig1}(g) we also mark the Fermi surface region we measure in main text Figs. 2(a-e) (white dotted line) and the Fermi surface region presented below in Fig. \ref{Fig2} (green dotted line). From \textit{ab initio} calculation, we find that \comp\ has 8 Weyl points, all above the Fermi level, all on the $k_z = 0$ plane of the Brillouin zone and all roughly located at $k_y \sim \pm k_{\textrm{W}}$, as shown schematically in Fig. \ref{Fig1}(h,i). We can understand the Weyl points as arising from valence and conduction bands which approach each other more or less tangentially, forming pairs of Type II Weyl points. 

\begin{figure*}[h!]
\centering
\includegraphics[width=15cm]{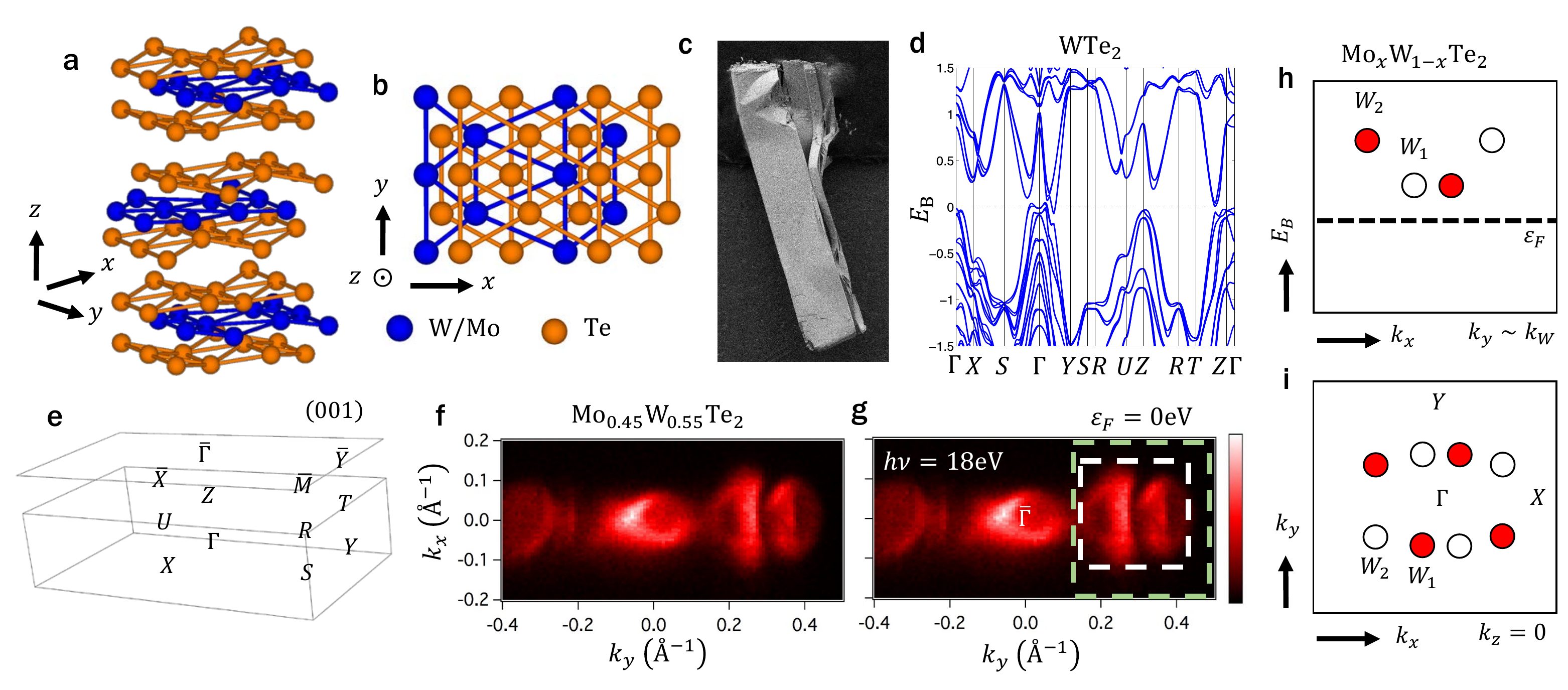}
\caption{\label{Fig1} \textbf{Overview of \comp}. (a) Side view of the \comp\ crystal structure. \comp\ crystallizes in an orthorhombic structure with space group $Pmn2_1$ ($\#31$). The W/Mo atoms are shown in blue and the Te atoms in orange. (b) Top view of \comp, showing only one monolayer, consisting of a layer of W/Mo atoms sandwiched between two layers of Te atoms. The structure lacks inversion symmetry. (c) A beautiful scanning electron microscope (SEM) image of a $\sim{0.4 \times 2}$ mm rectangular shaped \half\ bulk crystal clearly demonstrates a layered structure. (d) The electronic bulk band structure of WTe$_2$. There is a band inversion near $\Gamma$ which gives rise to electron and hole pocket along the $\Gamma$-$Y$ direction. (e) The bulk and (001) surface Brillouin zone (BZ) of \comp\ with high symmetry points labeled. (f,g) Fermi surface by vacuum ultraviolet APRES on the (001) surface of \half\ at $\varepsilon_{F}=0$eV with an incident photon energy of $18$ eV. We observe a palmier pocket and an almond pocket on either side of $\bar{\Gamma}$ along the $\bar{\Gamma}$-$\bar{Y}$ direction. The pockets correspond well to the electron and hole pockets we find in our \textit{ab initio} calculations, as presented in main text Fig. 2(l-n). We indicate the Fermi surface region presented in main text Figs. 2(a-e) (white dotted line) and the region presented below (green dotted line). We expect to see Weyl points and Fermi arcs above the Fermi level in the region where the electron and hole pockets intersect. (h,i) Distribution of Weyl points from \textit{ab initio} calculation in \comp. All Weyl points lie above the Fermi level, they are all Type II, they lie on $k_z = 0$ and roughly on the same $k_y \sim \pm k_\textrm{W}$. It is clear that to directly observe the Weyl points or calculate Chern numbers in \comp, it is necessary to access the unoccupied band structure.}
\end{figure*}

\bigskip
{\bf $\S 2.$ Systematic ARPES data on \half}
\bigskip

We present an additional Fermi surface mapping at low photon energy to complement the mapping presented in main text Fig. 2(a-e). Specifically, in Figs. \ref{Fig2}(a-c), we present a mapping at $h\nu = 6.36$ eV, which highlights the entire palmier and almond, unlike the mapping at $h\nu = 6.05$ eV, which provides beautiful contrast where the two pockets approach each other. We clearly see that the palmier is hole-like, while the almond is electron-like, supporting our discussion in the main text. We label the electron and hole pockets and two $E_\textrm{B}-k_x$ cuts in Fig. \ref{Fig2}(d). In Figs. \ref{Fig2}(e,f) we see that the electron and hole pockets closely approach each other, as seen also at $h\nu = 6.05$ eV and discussed in the main text. In Fig. \ref{Fig2}(g) we mark the electron and hole pockets, emphasizing that the hole pocket has a concave shape so that it nests the electron pocket.

\begin{figure*}[h]
\centering
\includegraphics[width=15cm]{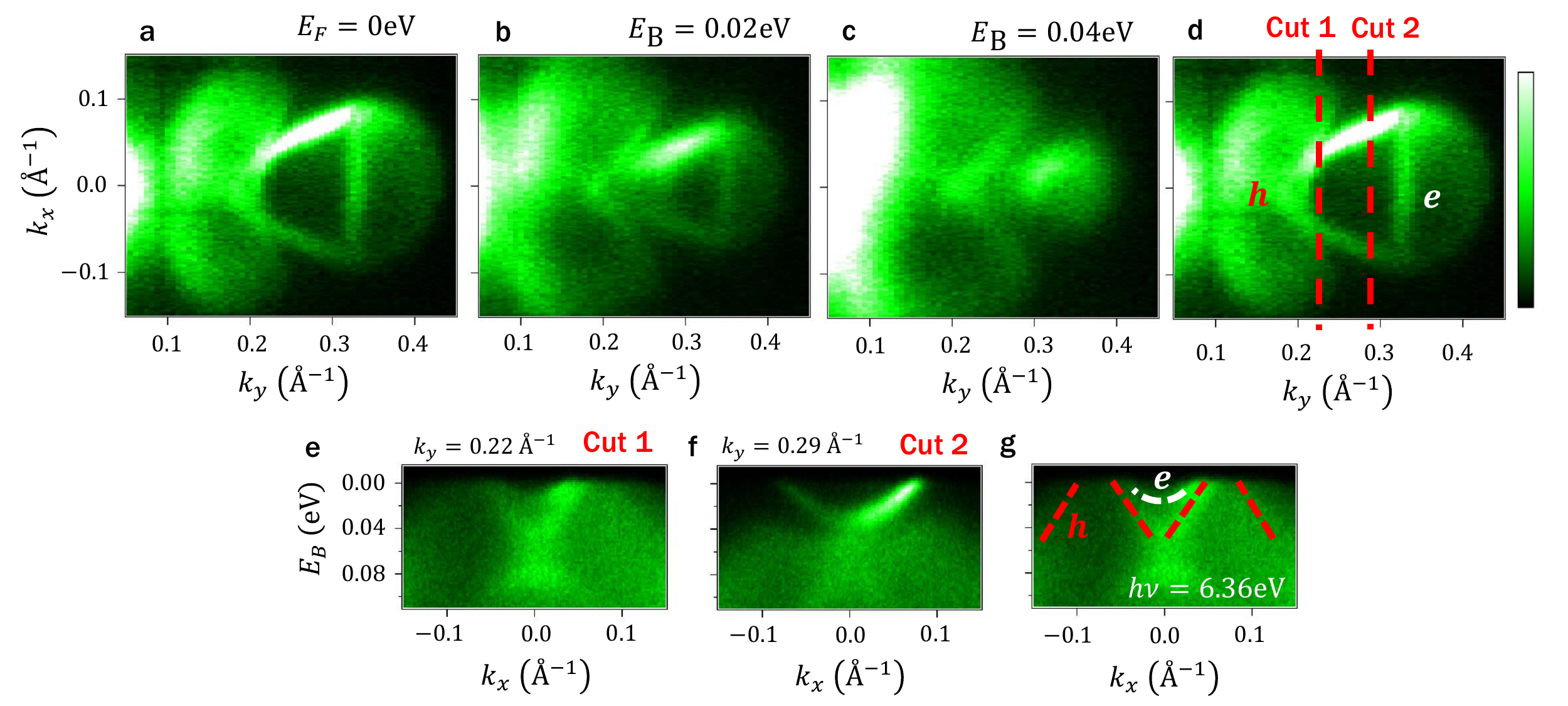}
\caption{\label{Fig2} \textbf{Occupied band structure of \comp}. (a-c) Fermi surface at different binding energies, clearing showing a hole-like palmier pocket and an electron-like almond pocket. (d) Same as (a), with the hole and electron pockets labeled, and with the cuts shown below marked. (e,f) Cuts near $k_y \sim k_\textrm{W}$ showing the electron and hole bands approach close to one another. We expect the electron and hole pockets to intersect above the Fermi level at $k_y \sim k_\textrm{W}$, forming Weyl points and Fermi arcs, see main text. (g) Same as (e), but with the electron and hole pockets marked. The hole pocket is partly concave, allowing it to hug the electron pocket.}
\end{figure*}

\bigskip
{\bf $\S 3.$ Methods: ARPES measurements}
\bigskip

Synchrotron-based ARPES measurements were performed at the CASSIOPEE beamline at Soleil in Saint-Aubin, France and the I05 beamline at the Diamond Light Source (DLS) in Didcot, United Kindom. ARPES measurements were also carried out using a home-built laser-based ARPES setup at the Ames Laboratory in Ames, Iowa, United States. All measurements were conducted under ultra-high vacuum and at temperatures $\leq 10$K. The angular and energy resolution of the synchrotron-based ARPES measurements was better than $0.2^{\circ}$ and $20$ meV, respectively. Photon energies from 15 eV to 100 eV were used. The angular and energy resolution of the home-built laser-based ARPES measurements was better than $0.1^{\circ}$ and $5$ meV, respectively, with photon energies from 5.77 eV to 6.67 eV \cite{KaminskiMethods}.

The pump-probe ARPES apparatus consisted of a hemispherical analyzer and a mode-locked Ti:Sapphire laser system that delivered $1.48$ eV pump and $5.92$ eV probe pulses at a $250$kHz repetition \cite{IshidaMethods}. The time and energy resolution was $300$ fs and $27$ meV, respectively. The spot diameter of the pump and probe beams at the sample position was $250$ and $85 \mu$m, respectively. Samples were cleaved in the spectrometer at $<5 \times 10^{-11}$ Torr, and measurements were conducted at $\sim8$ K.

\bigskip
{\bf $\S 4.$ Methods: \half\ growth}
\bigskip

High quality ribbon-like single crystals of Mo$_x$W$_{1-x}$Te$_2$ were grown by chemical vapor transport (CVT) with iodine (I) as the agent. Before growing the crystals, the quartz tubes were thoroughly cleaned using thermal and ultrasonic cleaning treatments to avoid contamination. Stoichiometric amounts of W (99.9\% powder, Sigma-Aldrich), Mo (99.95\% powder, Sigma-Aldrich) and Te (99.95\%, Sigma-Aldrich) were mixed with iodine, were sealed in a $20$ cm long quartz tube under vacuum $\sim 10^{?6}$ Torr, and then placed in a three-zone furnace. The reaction zone dwelled at $850$ $^{\circ}$C for 40 hours with the growth zone at $900$ $^{\circ}$C and then heated to $1070$ $^{\circ}$C for seven days with the growth zone at $950$ $^{\circ}$C. Lastly, the furnace was allowed to cool naturally down to room temperatures. The Mo$_x$W$_{1-x}$Te$_2$ single crystals were collected from the growth zone. Excess iodine adhering to the single crystals was removed by using acetone or ethanol. Nominal compositions of the Mo$_x$W$_{1-x}$Te$_2$ crystals were $x = 0.0, 0.1, 0.15, 0.2, 0.3$. An energy dispersive spectroscopy (EDS) measurement was carried out to precisely determine the composition of the samples. Samples were first surveyed by an FEI Quanta 200FEG environmental scanning electron microscope (SEM). The chemical compositions of the samples were characterized by an Oxford X-Max energy dispersive spectrometer that was attached to the SEM. All the samples were loaded at once in the SEM chamber to ensure a uniform characterization condition. Both the SEM imaging and EDS characterization were carried out at an electron acceleration voltage of $10$ kV with a beam current of $0.5$ nA. For each sample, three different spatial positions were randomly picked to check the uniformity. The measured compositions were: MoTe$_2$ for $x = 0$, Mo$_{0.42}$W$_{0.58}$Te$_2$ for $x= 0.1$,  Mo$_{0.46}$W$_{0.54}$Te$_2$ for $x = 0.15$, Mo$_{0.44}$W$_{0.56}$Te$_2$ for $x = 0.2$ and Mo$_{0.43}$W$_{0.57}$Te$_2$ for $x = 0.3$. For $x \neq 0$, we considered the samples to all have the approximate composition Mo$_{0.45}$W$_{0.55}$Te$_2$.

\bigskip
{\bf $\S 5.$ Methods: single-crystal XRD}
\bigskip

A black, bar-shaped crystal with approximate dimensions $0.11 \times 0.09 \times 0.20$ mm on the tip of a glass fiber was selected for data collection. The single crystal diffraction data were collected on a SuperNova X-ray diffraction system from Agilent Technologies equipped with a graphite-monochromated MoK$\alpha$ radiation ($\lambda = 0.71073 \AA$) at room temperature. The data were corrected for Lorentz factors, polarization, air absorption, and absorption due to variations in the path length through the detector faceplate. Absorption correction based on a multi-scan technique was also applied. Absorption corrections were performed by the SADABS program \cite{XRD1,XRD2}. The space group was determined to be $Pmn2_1$ ($\#31$) based on systematic absences, E-value statistics, and subsequent successful refinements of the crystal structure. The structure was solved by the direct method and refined by full-matrix least-squares fitting on $F^2$ by SHELX-97. \cite{XRD3} There are two metal sites (M1 and M2) which are all octahedral coordination, and four sites for tellurium atoms. The M1 and M2 sites with mixed occupancy by W and Mo were refined and the formula was W$_{0.69}$Mo$_{0.31}$Te$_2$ where the occupancy for W and Mo in M1 and M2 sites were $0.679(4)$, $0.321(4)$ and $0.705(4)$, $0.295(4)$, whose charge was neutral. All atoms were refined anisotropically ($R_1 = 0.0469$, $wR_2 = 0.1006$). See Table \ref{tab1} for crystallographic data and structural refinements.

\begin{table}[]
\centering
\begin{tabular}{| c | c |} 
\hline
Formula & \ \half\ \ \\ \hline
\textit{fw} & 411.97 \\ \hline
Crystal system & orthorhombic \\ \hline
Crystal color & black \\ \hline
Space group & $Pmn2_1$ ($\#31$) \\ \hline
$a$ $(\textrm{\AA})$ & 3.4834(6) \\ \hline
$b$ $(\textrm{\AA})$ & 6.3042(10) \\ \hline
$c$ $(\textrm{\AA})$ & 13.9152(18) \\ \hline
$\alpha = \beta = \gamma$ (deg.) & 90 \\ \hline
$V$ $(\textrm{\AA}^3)$ & 293.79(4) \\ \hline
$Z$ & 4 \\ \hline
$D_c$ $(\textrm{gcm}^{-3})$ & 8.95418 \\ \hline
GOOF on $F^2$ & 1.002 \\ \hline
Flack	 & 0.00(4) \\ \hline
$R_1$, $wR_2 (I > 2\sigma(I))$ & 0.0402, 0.0954 \\ \hline
$R_1$, $wR_2 (\textrm{all data})$ & 0.0469, 0.1006 \\ \hline
Largest diff peak/hole, $e\textrm{\AA}^{-3}$ & $2.65 / -2.95$ \\ \hline
\end{tabular}
\caption{Crystal data and structure refinements from single-crystal X-ray diffraction of \half.}
\label{tab1}
\end{table}

\bigskip
{\bf $\S 6.$ Methods: \textit{ab initio} calculations}
\bigskip

We computed the electronic structures by using the projector augmented wave method \cite{PAW1,PAW2} as implemented in the VASP \cite{TransitionMetals, PlaneWaves1, PlaneWaves2, GGA} package within the generalized gradient approximation (GGA) schemes \cite{GGA}. For WTe$_2$, the experimental lattice constants used were from \cite{MoTe2WTe2}. A $8 \times 16 \times 4$ Monkhorst Pack $k$-point mesh was used in the computations. The lattice constants and atomic positions of MoTe$_2$ were fully optimized in a self-consistent calculation for an orthorhombic crystal structure until the force became less than $0.001$ eV$/\textrm{\AA}$. Spin-orbit coupling was included in our calculations. To calculate the bulk and surface electronic structures, we constructed a first-principles tight-binding model Hamiltonian for both WTe$_2$ and MoTe$_2$, where the tight-binding model matrix elements were calculated by projecting onto the Wannier orbitals \cite{MLWF1,MLWF2,Wannier90}, which used the VASP2WANNIER90 interface \cite{MLWF3}.  We used the $s$- and $d$-orbitals for W(Mo) and the $p$-orbitals for Te to construct Wannier functions, without performing the procedure for maximizing localization. The electronic structure of Mo$_x$W$_{1-x}$Te$_2$ samples was calculated by a linear interpolation of the tight-binding model matrix elements of WTe$_2$ and MoTe$_2$. The surface states were calculated by the surface Green's function technique \cite{Green}, which computed the spectral weight near the surface of a semi-infinite system.

\end{document}